\documentclass[conference]{IEEEtran}
\IEEEoverridecommandlockouts
\usepackage{cite}
\usepackage{amsmath,amssymb,amsfonts,url}
\usepackage{algorithmic}
\usepackage{graphicx}
\usepackage{textcomp}
\usepackage{xcolor}
\def\BibTeX{{\rm B\kern-.05em{\sc i\kern-.025em b}\kern-.08em
    T\kern-.1667em\lower.7ex\hbox{E}\kern-.125emX}}
\begin{document}

\title{Classification of Single-lead Electrocardiograms: TDA Informed Machine Learning\\
}
\author{\IEEEauthorblockN{Christopher Dunstan}
\IEEEauthorblockA{\textit{University of Maryland Baltimore County} \\
Baltimore, MD, USA\\
cdun2@umbc.edu}
\and
\IEEEauthorblockN{Esteban Escobar}
\IEEEauthorblockA{\textit{California Polytechnic State University Pomona} \\
Pomona, CA, USA \\
estebane@cpp.edu.edu}
\and
\IEEEauthorblockN{Paul Samuel Ignacio}
\IEEEauthorblockA{\textit{University of the Philippines Baguio} \\
Baguio City, Philippines \\
ppignacio@up.edu.ph}
\and
\IEEEauthorblockN{Luke Trujillo}
\IEEEauthorblockA{\textit{Harvey Mudd College} \\
Claremont, CA, USA \\
ltrujillo@g.hmc.edu}
\and
\IEEEauthorblockN{David Uminsky}
\IEEEauthorblockA{\textit{University of San Francisco} \\
San Francisco, CA, USA \\
duminsky@usfca.edu}
}

\maketitle

\begin{abstract}
Atrial Fibrillation is a heart condition characterized by erratic heart rhythms caused by chaotic propagation of electrical impulses in the atria, leading to numerous health complications. State-of-the-art models employ complex algorithms that extract expert-informed features to improve diagnosis. In this note, we demonstrate how topological features can be used to help accurately classify single lead electrocardiograms. Via delay embeddings, we map electrocardiograms onto high-dimensional point-clouds that convert periodic signals to algebraically computable topological signatures. We derive features from persistent signatures, input them to a simple machine learning algorithm, and benchmark its performance against winning entries in the 2017 Physionet Computing in Cardiology Challenge. 
\end{abstract}


\section{Introduction}
Cardiac arrhythmia, or abnormal heart rhythm, is the most prevalent heart disorder that encompasses a wide array of conditions from heart rate abnormalities (like Bradycardia and Tachycardia), to premature heartbeats, and erratic rhythms. Among these, Atrial Fibrillation (AFib) is the most common, affecting 33.5 million people worldwide in 2010 (Chugh et al. \cite{chugh}). AFib is characterized by erratic heart rhythms caused by chaotic propagation of electrical impulses in the atria. This triggers atrial spasms and irregular opening and premature closing of the atrioventricular valves, resulting in an increased risk of clot formation, and in the extreme case, stroke.

An electrocardiogram (ECG or EKG) is the main tool that medical professionals use to diagnose AFib, measuring electric activity in the heart at different stages of the cardiac cycle. Central to the analysis of ECG measurements is the PQRST complex, an important slice of an ECG reading composed of a series of wave patterns that mark specific events in the cardiac rhythm. There is much existing literature on the analysis of specific features in the PQRST complex and other parts of a standard 12-lead ECG recording that help in the diagnosis of many heart conditions. In particular, state-of-the-art models employ advanced algorithms that extract expert-informed features from the PQRST complex to diagnose AFib, and a large majority of these focus on two features: the P waves and the RR intervals. P waves record atrial depolarization and correspond to electrical activity in the atria prior to transfer of blood to the ventricles, whereas RR intervals measure the time between the peaks of ventricular depolarization, the cardiac event corresponding to the ventricles pumping the blood out of the heart. Because electrical impulses in the atria are in chaos, clear P waves tend to be absent in ECG readings of people with AFib, and ventricular activity become irregular causing the RR intervals to be highly variable. These two features demonstrate how specific (i.e. local) landmarks in the structure of an ECG reading can be used to examine tell-tale signs of abnormal cardiac activity.  

In this paper, we explore whether the local and global topology of ECG readings, paired with minimal medical knowledge, could be utilized to aid in the diagnosis of AFib. Our goal is to provide experimental answer to the following question: \emph{Is there signal in the topological features of ECG readings in diagnosing Atrial Fibrillation?} 
This query is important as it explores the call for increased collaboration between theoretical focus and advances of topologists and the medical researchers.

\subsection{Pipeline}
Our approach consists of four stages: ECG cleanup, point cloud generation, topological feature extraction, and classification via random forest. 
\begin{figure}[h]
\label{pipeline}
\centering
\includegraphics[width=0.45\textwidth]{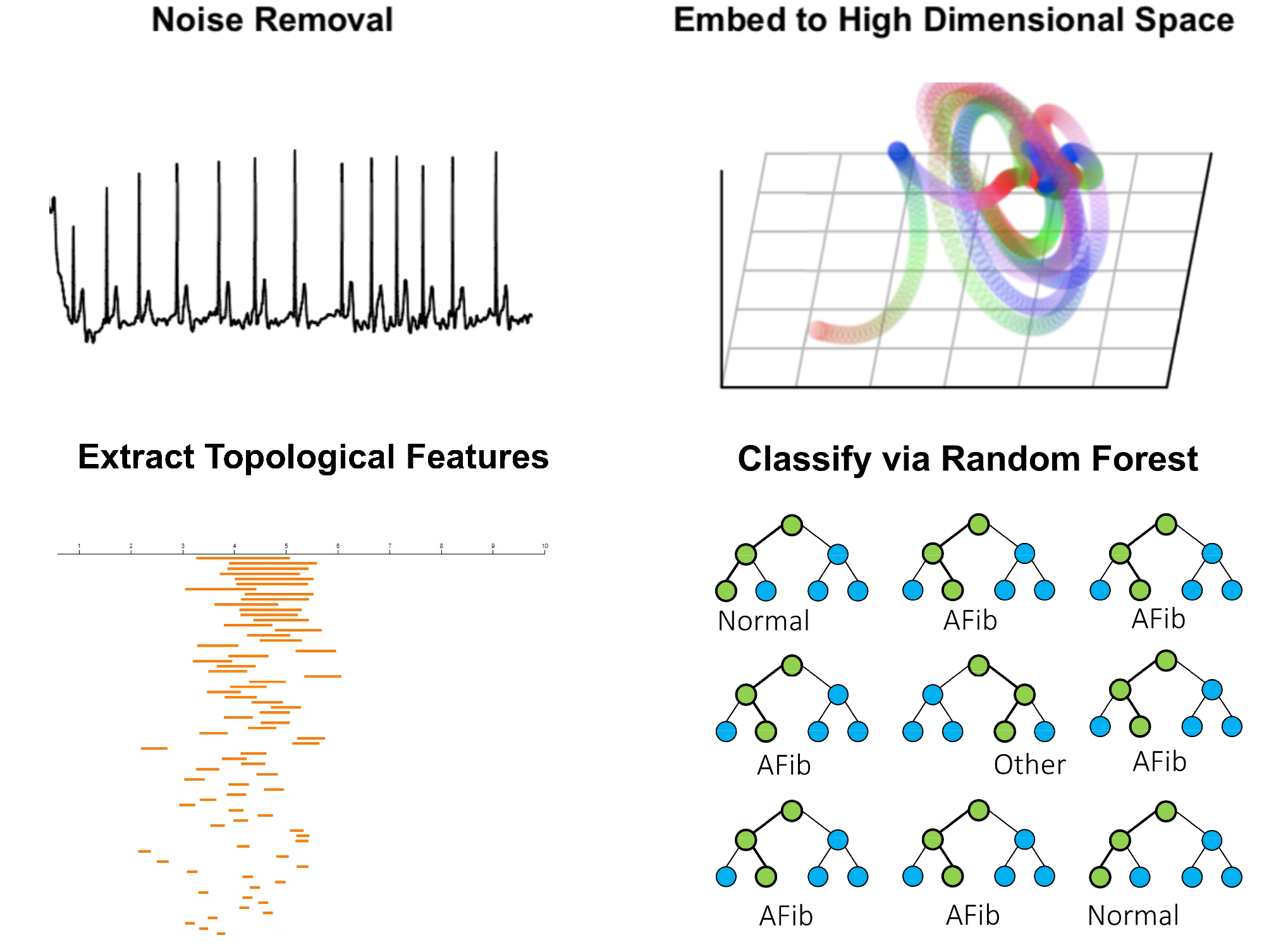}
\caption{The pipeline of our topology-based classification of ECG readings.}
\end{figure}

Integral to our objective of detecting signal from the topology of ECG readings as time series is to mitigate the effects of measurement error and noise. We implement a simple signal frequency-based approach in extracting the longest and ``cleanest'' portion of an ECG recording. We then transform each time series as a point cloud in high-dimensional space and examine its topological features using novel tools from topological data analysis. This step assigns to each point cloud a distance-parameterized summary of its evolving topological features. Finally, we derive statistical features from this summary to train a simple random forest classifier. 

It is worth noting that the idea of studying the topology of high-dimensional point-clouds as embeddings of time series data is fairly recent and has been explored in several applications. The basic difference among these applications is on the treatment of the computed features. Perea et al. \cite{perea1} used the most significant 1-dimensional topological signature from the summary to score periodicity in gene expression time series data. Seversky et al. \cite{seversky} also followed the same pipeline, but used metrics and kernels on the space of topological summaries to study classification of time series data. Finally, Umeda \cite{umeda} introduced a variation of the output topological summary compatible as input to Convolutional Neural Networks for classifying volatile time series.

\subsection{The Data}

In 2017, Physionet and Computing in Cardiology launched a challenge to develop algorithms able to classify single lead electrocardiogram readings ranging in length from 9 to 60 seconds into four categories: normal sinus rhythm, AFib, other sinus rhythm, or noisy. A total of 12,186 electrocardiogram recordings, donated by AliveCor (\url{www.alivecor.com}), split into a training set (8,528) and a hidden test set (3,658) were used in the challenge. Initial labeling of the released training set follows a distribution of 60.4\% normal, 9\% AFib, 30\% other rhythms, and 0.5\% noisy. This distribution is later revised to 59.5\% normal, 8.9\% AFib, 28.3\% other rhythms, and 3.3\% noisy following a re-labeling step due to inter-expert disagreement on a significant fraction of the labels --- a testament to the challenge's difficulty and the existing disagreements in real practice. A comprehensive report on the challenge is provided by the organizers (Clifford et al. \cite{clifford}). We utilize the data set used in this challenge with the final labeling information.
\section{Methods}
\subsection{Noise Removal}
We adapt the strategy for noise clean up from Datta et al. \cite{datta} to our approach. The spectogram of each ECG reading is computed and portions along the time axis with spectral power above 50 Hz are cut-off. This produces segments of the original ECG reading sanitized from extreme noise (literature pegs important cardiac information to be within 20 Hz) caused by measurement irregularities. We use the first 3000 time points of the longest clean segment when possible, otherwise we use the first 3000 time points of the original ECG reading.

\subsection{Sliding Window Embeddings}
The method of converting time series data into point clouds via \emph{sliding window embedding} (or \emph{delay embedding}) has been explored in many types of applications. The general idea of capturing rich local information within slices (i.e. windows) of the time series, and recording them as vectors in high-dimensional space circumvents many issues that come with sampling within the time series given its discrete nature. It also illuminates global structures of time series as artifacts of the local dynamics, a powerful consequence of the famous embedding theorems of Whitney \cite{whitney} and Takens \cite{takens}, provided parameters are chosen appropriately. Furthermore, it has been shown that this technique increases precision of parameter estimates for modeling variability in recurring phenomenon for time-dependent data (von Oertzen and Boker \cite{von}).

The embedding process begins by selecting a window size $w$ and embedding dimension $d$. These two respectively control the scope and resolution at which local dynamics will be observed. A window of length $w$ corresponding to the starting $w$ time units of the time series is first considered from which $d$ time point measurements are extracted. These $d$ measurements together define a vector in $\mathbb{R}^d$, and is the first element in the point-cloud embedding. The window then slides at step size $\tau$, and the process is repeated, mapping the next window to another vector in the high-dimensional space (see first time series in Figure 2). One advantage of this approach is that the topological features of the embedded space remain invariant under inversion, i.e. flipping upside down (see third time series in Figure 2), of the time series, bypassing the problem of identifying whether or not an ECG reading is inverted --- an existent issue in the data set.
\begin{figure}[h]
\label{embed}
\centering
\includegraphics[width=0.5\textwidth]{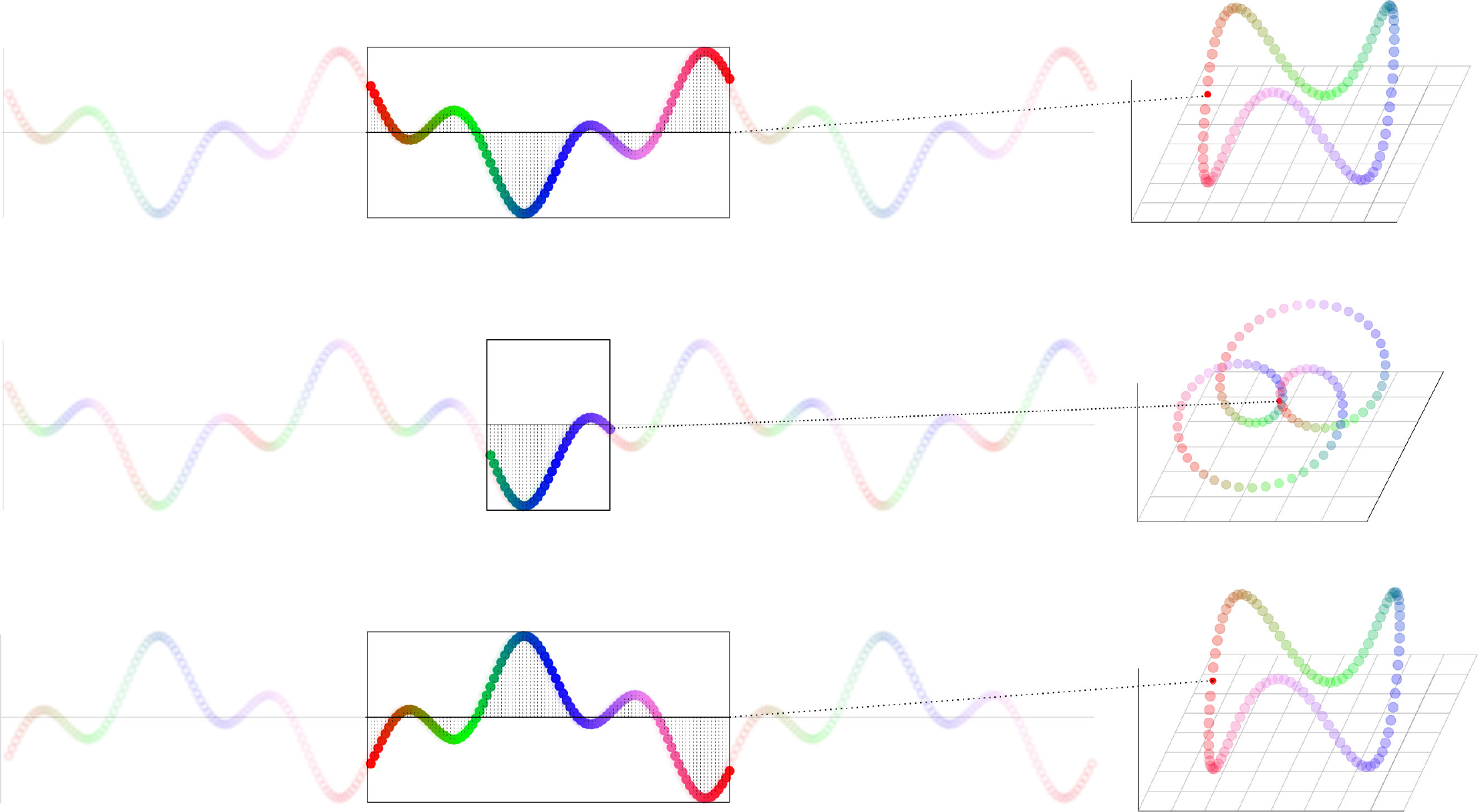}
\caption{Multi-dimensional scaling to $\mathbb{R}^3$ of the high-dimensional point clouds generated from delay embeddings with varying window sizes (respectively 100, 35, and 100 time units). Here, the embedding dimensions are set equal to the window sizes and the delay parameter $\tau$ is set to 1. Projection to $\mathbb{R}^3$ is only used for visualization.}
\end{figure}
It is clear that changing the window size and/or the embedding dimension would drastically alter the resulting embedded structure (see second time series in Figure 2), prompting a careful selection of these input parameters in our analysis of the ECG readings. We now discuss these choices.

An ECG reading is naturally periodic, mimicking the cardiac cycle. The embedding process converts the periodic patterns present in the time series to attractor cycles in the high-dimensional point-cloud. For ECG readings, this pattern pertains to the PQRST complex, suggesting that parameters must be chosen to capture local dynamics within and around it. Furthermore, in view of the succeeding stage in our approach, we also would like that the resulting point-cloud be as \emph{``round''} as possible to maximize the diameter of the resulting cycle attractors. This has been shown to hold when the window size is chosen to be as close as possible to the period of the pattern (Perea and Harer \cite{perea2}). After close inspection of the ECG readings, we determine that this is approximately 250 ms.

For the embedding dimension, we select the optimal choice based on computational efficiency and stability of the resulting topological summary, that is, we choose the embedding dimension producing a topological summary that is most similar to those with neighboring dimensions. We compare topological summaries using the bottleneck distance, a standard metric used in topological data analysis that measures the cost of tranforming one topological summary to another, and is central to the discussion on stability of the output summaries under slight perturbation of data \cite{stab}. Figure 3 shows the boxplots of the bottleneck distances between paired topological summaries from neighboring dimensions. We maintain a balance between selecting bottleneck differences that are not too spread while accounting for sparsity of points in the embedded point cloud since a dimension that is too low selects too few points from the window and too high produces very expensive computations. 
\begin{figure}[h]
\centering
\includegraphics[width=0.5\textwidth, height = 2in]{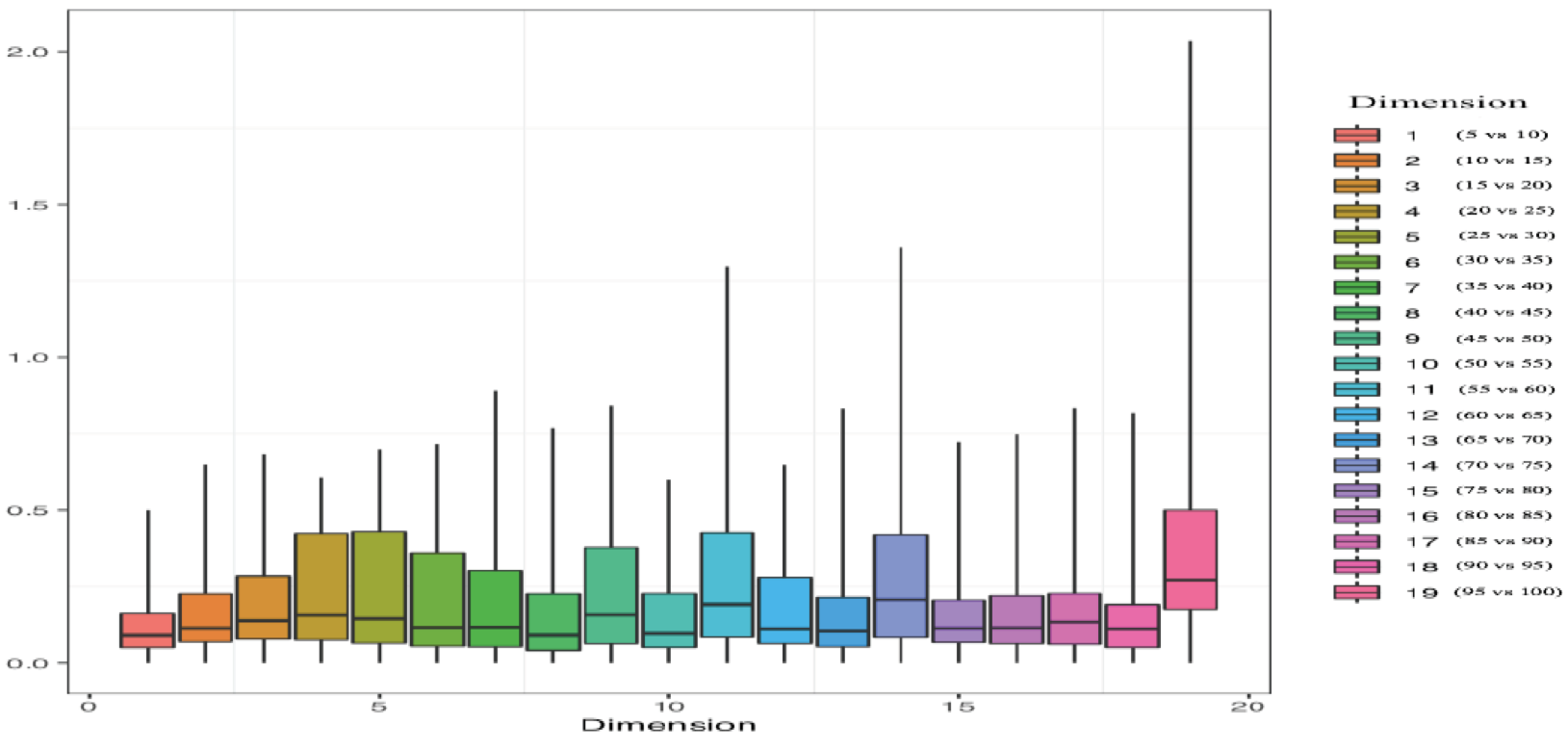}
\caption{Boxplots of bottleneck distances between paired barcodes from neighboring embedding dimensions. Examination of boxplot number 10 suggests an optimal embedding dimension of $50$.}
\end{figure}

\subsection{Feature Extraction via Persistent Homology}
To each point cloud, we apply a tool from topological data analysis known as \emph{persistent homology} to extract evolving topological features. This is a relatively new approach in data analysis that has been growing in popularity because of its novel treatment of data as topological objects, and has been applied to a wide array of data sets including images \cite{carlsson}, brain data \cite{hess,memoli}, migration data \cite{ignacio}, and recently, time series \cite{seversky, perea2}. In this section, we discuss the fundamental ideas of this approach, and provide insights as to the meaning of computable topological signatures in time series data. For a more in-depth introduction to persistent homology, we refer the interested reader to \cite{patania,otter,zomo}.

To start, given a fixed threshold $\varepsilon$, we endow the point cloud with a \emph{Vietoris-Rips Complex} structure by treating as $n$-dimensional objects a collection $\{p_0,p_1,...,p_n\}$ of \emph{n+1} points (called an \emph{n-simplex}) whenever $d(p_i,p_j)\leq \varepsilon$ for all pairs $0\leq i,j\leq n$, where $d$ is a defined metric in the ambient space (see \cite{ripsref1, otter} for a more detailed description of the Vietoris-Rips Complex). For our point cloud embeddings generated from sliding windows, the points live in $\mathbb{R}^{50}$ and the metric is the Euclidean distance. A way to visualize these objects is to consider a \emph{0-simplex} $\{p_i\}$ as a point, a \emph{1-simplex} $\{p_i,p_j\}$ as an edge through $p_i$ and $p_j$, a \emph{2-simplex} $\{p_i,p_j,p_k\}$ as a triangle having $p_i, p_j,$ and $p_k$ as vertices, and so on. This allows one to view the point cloud as a collection of mathematical pieces, called \emph{Vector Spaces}, $\Lambda_0, \Lambda_1,\Lambda_2,...,\Lambda_n,...$ where each $\Lambda_i$ is built up from the $i$-dimensional simplices and related by maps $\partial_n: \Lambda_n\to \Lambda_{n-1}$ sending a $n$-dimensional object $\sigma_n\in \Lambda_n$ to its \emph{boundary} $\partial_n(\sigma_n)\in \Lambda_{n-1}$. This construction further generates abstract algebraic objects, called \emph{homology groups}, whose signatures (\emph{Betti numbers}) $\beta_0, \beta_1,...$ encode topological information about the underlying point cloud: $\beta_0$ counts connected components, $\beta_1$ loops or holes, $\beta_2$ voids, and so on. These are the features that we are interested in. The reader may consult standard references in algebraic topology such as \cite{hatcher, munkres} for a thorough exposition on these ideas.

There is, however, one caveat: the signatures that homology captures depend on the simplicial structure constructed via a choice for the threshold $\varepsilon$. From this, a natural question arises: how must $\varepsilon$ be chosen? A solution that topological data analysis proposes circumvents this by instead considering a sequence of simplicial structures induced by increasing the threshold parameter $\varepsilon$, and keeping track of topological features that survive as $\varepsilon$ varies. This process then records the evolution of topological features of the point cloud and is the main idea of \emph{persistent homology}. The topological features detected by persistent homology are recorded in a summary called a \emph{persistence barcode} (see Figure 4), showing the lifetime of a detected feature and its relative significance with respect to all features detected --- in persistent homology, long bars represent significant features while short bars detect noise.
\begin{figure}[h]
\centering
\includegraphics[width=0.48\textwidth]{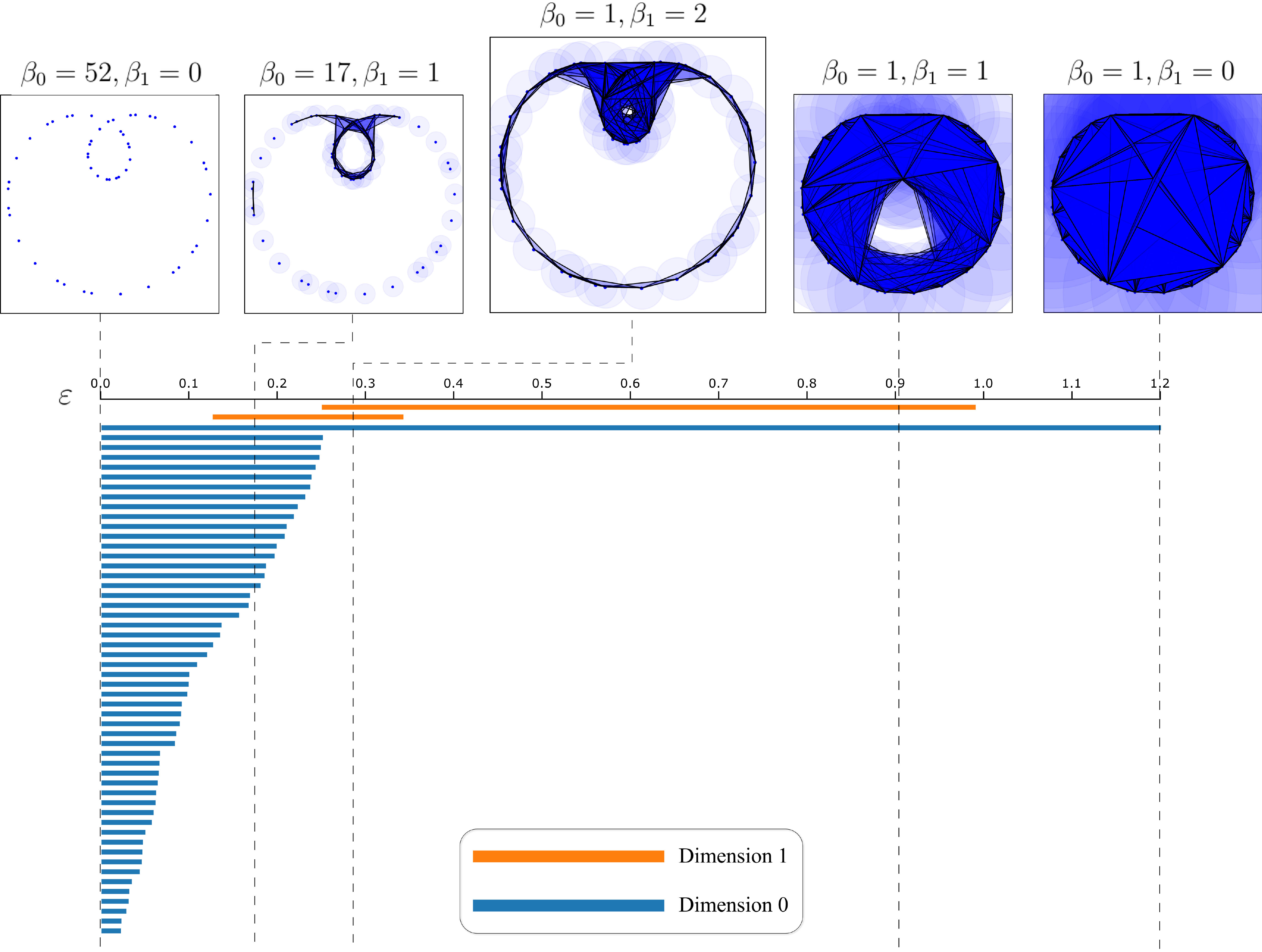}
\caption{The persistence barcode of a point cloud sampled from a cardioid superimposed with simplicial complexes and the correspoding Betti numbers at different thresholds.}
\end{figure}

For non-geometric data, it can be a challenge to interpret what kind of information topological features reveal. However, in our setting, these features have clear meaning. For the 1-dimensional case, the features represent cycles in the point cloud induced from sliding window embeddings. Since cycle attractors in this point cloud correspond to periodic patterns in the corresponding time series, the 1-dimensional features detect periodic information about the original time series. This observation is the basis for the \emph{SW1PerS} algorithm proposed by Perea et al. \cite{perea1} for scoring periodicity in time series data.

To examine if these topological features contain signal for ECG diagnosis, we derive simple statistical summaries from the features based on the barcode of each ECG. Table \ref{statfeatures} summarizes those that are found to improve the accuracy of the random forest when included in the feature set. The summary measures in this table follow the standard definitions in Statistics: \emph{Mean} refers to the arithmetic average, \emph{Standard Deviation (SD)} measures the spread of the values around the mean,  \emph{Skewness} quantifies the symmetry or asymmetry of a set of values, \emph{Kurtosis} measures the weight of the tails of the distribution relative to the center, and \emph{Sum} refers to the total. 

\begin{table}[h]
\begin{center}
\caption{\label{statfeatures} Statistical Features from Barcodes}
\footnotesize
\begin{tabular}{@{}lccccc}
\textbf{Barcode Feature}			&\textbf{Mean}	&\textbf{SD}	&\textbf{Skewness}	&\textbf{Kurtosis} & \textbf{Sum}\\
\hline
\textit{Dimension 0}				&			& 			&		&	&\\
\hline
\hspace{0.3cm} Death			&\checkmark 	& 			& \checkmark				&\checkmark&	\\
\hline
\textit{Dimension 0}$^{\mathrm{*}}$ 	&			& 			& 				&	&\\
\hline
\hspace{0.3cm} Death			&\checkmark	&\checkmark 	&\checkmark		&	&\\
\hline
\textit{Dimension 1}				&			& 			& 				&	&\\
\hline
\hspace{0.3cm} Birth 			& 			&\checkmark	&\checkmark		&	&\\
\hline
\hspace{0.3cm} Death			& 			& 			& 				&\checkmark	&\\
\hline
\hspace{0.3cm} Persistence		& \checkmark	& 			& 				&	&\checkmark\\
\hline
\textit{Dimension 1}$^{\mathrm{*}}$ 	&			& 			& 				&	&\\
\hline
\hspace{0.3cm} Persistence		& 			& 			& 				&\checkmark	&\checkmark\\
\hline
\textit{Dimension 2}				&			& 			& 				&	&\\
\hline
\hspace{0.3cm} Birth 			& 			& 			& 				&\checkmark	&\\
\hline
\hspace{0.3cm} Death			&\checkmark	& 			& 				&	&\\
\hline
\textit{Dimension 2}$^{\mathrm{*}}$ 	&			& 			& 				&	&\\
\hline
\hspace{0.3cm} Birth 			& 			& 			& \checkmark				&	&\\
\hline
\multicolumn{5}{l}{$^{\mathrm{*}}$These bars do not include the 5\% most persistent bars.}
\end{tabular}\\
\end{center}
\end{table}
In addition to these features, three others were included namely, the ratio of the length of the longest clean segment of the ECG with respect to the original length, and the mean and standard deviation of dimension 0 persistence obtained by filtering the ECG time series, considered as functions, via super-level sets. 
\subsection{Random Forest}
Once the feature set is extracted using persistent homology, we input this set into a random forest, an ensemble of decision trees each using a randomized set of features to decide the classification of an object. The idea is that if topological features are preserved within ECGs of the same type, then the random forest will learn about these intra-class descriptors and use these as basis to provide a good classification for a previously unseen ECG.
 
To examine if these topological features contain useful information for ECG diagnosis, we set up two random forest models. Since the hidden test set from the Physionet Challenge was never released, we extract a test set of 1000 ECG readings proportionally chosen randomly within each class to reproduce the same distribution as the competition test set. We then bootstrap the remaining ECGs for training to recover a training set of comparable size to the original competition training set. The first model is trained using four statistical features based on the RR intervals --- a known good differentiator of ECG readings. In addition to these four features, the second model includes the other statistical features derived from the barcodes. Both models are given the same training and test set. We perform this approach 100 times, each time changing both the training and test set and recording the classification scores based on the Physionet Challenge metric
$$F_1 = (F_{1a}+F_{1n} + F_{1o})/3$$
where each of the scores $F_{1a}, F_{1n},$ and $F_{1o}$ are computed using the formula
\begin{align*}
F_{1a} &= 2Aa/(\Sigma A + \Sigma a),\\
F_{1n} &= 2Nn/(\Sigma N + \Sigma n),\\
F_{1o} &= 2Oo/(\Sigma O + \Sigma o)
\end{align*}
according to the table below:
\begin{table}[h]
\begin{center}
\caption{Table of values used for the $F_1$ score formula}
\begin{tabular}{cccccc}
&\multicolumn{5}{c}{\textbf{Predicted Classification}} \\
 & \textit{AFib}& \textit{Normal}& \textit{Others} & \textit{Noisy} & \textit{Total} \\
\textit{Afib} &$Aa$&$An$&$Ao$&$Ap$&$\Sigma A$\\
\textit{Normal} &$Na$&$Nn$&$No$&$Np$&$\Sigma N$\\
\textit{Others} &$Oa$&$On$&$Oo$&$Op$&$\Sigma O$\\
\textit{Noisy} &$Pa$&$Pn$&$Po$&$Pp$&$\Sigma P$\\
\textit{Total} &$\Sigma a$&$\Sigma n$&$\Sigma o$&$\Sigma p$&\\
\end{tabular}
\label{tab1}
\end{center}
\end{table}

Finally, a paired $t$-test is performed to examine if the differences in classification scores when topological features are included in the feature set are significant (as opposed to just the four basic RR interval features).

\section{Results and Discussions}
Table \ref{FinalF} shows the final $F_1$ scores of the two random forest models. For comparison, we include the scores of the winning models in the Physionet Challenge. The scores for validation come from validation set of 300 ECG readings  prepared by the Physionet Challenge. It must be pointed out that the test set used for our random forest models are only about one-third in size of the hidden test set from the challenge but has the same distribution. Moreover, since our test set is set aside from the training set, it also means that the training set used by the forests are reduced in size of distinct ECG readings (across classes) by the same amount. 

\begin{table}[h]\label{FinalF}
\caption{Final $F$ Scores of Different Models}
\begin{center}
\begin{tabular}{|c|c|c|c|c|}
\hline
Model 	&No. of Features	&Train	&Validation	&Test$^{\mathrm{*}}$ \\
\hline
Teijeirio et al. 	&86	&0.893	&0.912	&0.831 \\
\hline
Datta et al. 	&150	&0.970	&0.990	&0.829 \\
\hline
Zabihi et al. 	&150	&0.951	&0.968	&0.826 \\
\hline
Hong et al. 	&622	&0.970	&0.990	&0.825 \\
\hline
RF w. RR Features 	&4	&0.926	&0.920	&0.684 \\
\hline
RF w. RR \& TDA 	&23	&0.997	&0.975	&0.722 \\
\hline
\end{tabular}
$^{\mathrm{*}}$The test sets used by the two random forest models in each of the 100 cycles of training and testing are comparable to each other but not with the hidden test set used by the first four models from the Physionet Challenge. 
\label{tab1}
\end{center}
\end{table}

We highlight that with just twenty-three features, most of which are statistics from the persistence barcodes, the random forest model already performs relatively well with respect to the winning models. It is worth noting that all the winning models from the Physionet Challenge used features based on the RR intervals, and that just the four statistical features from the RR intervals already account for a significant portion of the $F_1$ scores. In addition, most of the features used by the winning highly tuned models (some include deep learning algorithms) are engineered based on features known to be helpful in diagnosing AFib and either employ advanced algorithms for extraction or medical expertise for processing. On the other hand, we purposely did not fine tune our model as we wanted to focus on whether or not there was any noticeable increase using TDA-based features.

Table IV provides a more detailed summary of the random forest model's performance across classes. Here, we see that features from the RR intervals are the main drivers of accuracy for all models, and that topology-based features consistently increase the accuracy across classes. More importantly, performing a paired one-tailed $t$-test at $\alpha=0.05$ between the class $F_1$ scores of the two random forests reveals that these increases in $F_1$ scores across classes are significant (see Table V). 

\begin{table}[h]\label{Ind}
\caption{$F_1$ Class Scores of Different Models}
\begin{center}
\begin{tabular}{|c|c|c|c|}
\hline
Model 	&$F_{1a}$	&$F_{1n}$	&$F_{1o}$\\
\hline
Teijeirio et al. 	&0.854	&0.903	&0.737	\\
\hline
Datta et al. 	&0.823	&0.916	&0.750	\\
\hline
Zabihi et al. 	&0.835	&0.909	&0.734	 \\
\hline
Hong et al. 	&0.823	&0.912	&0.751	\\
\hline
RF w. RR Features 	&0.649	&0.867	&0.536	 \\
\hline
RF w. RR \& TDA 	&0.688	&0.896	&0.580	 \\
\hline
\end{tabular}
\label{tab1}
\end{center}
\end{table}

\begin{table}[h]\label{Ttest}
\caption{Significant Increase in $F_1$ Scores Between the two Random Forest Models}
\begin{center}
\begin{tabular}{|c|c|c|c|c|c|}
\hline
	&$F_{1a}$	&$F_{1n}$	&$F_{1o}$ &$F_{1p}$ &$F_{1}$ \\
\hline
Significant Increase (\%)	&3.46	&2.69	&3.87 	&8.6	 &3.43 \\
\hline
$p$ value	&0.049	&0.042	&0.049 	&0.048	 &0.044 \\
\hline
\end{tabular}
\label{tab1}
\end{center}
\end{table}

\section*{Acknowledgment}

We are grateful to the Mathematical Sciences Research Institute for providing the best environment for research collaboration. We also would like to acknowledge the NSF (DMS-1659138), the NSA (H98230-18-1-0008), and the Sloan Foundation (G-2017-9876) for providing the grants that allowed us to complete the project. Author DU was partially supported by the Wicklow AI and Medical Research Initiative (WAMRI).

\end{document}